\documentstyle[aps,prb,twocolumn,graphicx]{revtex}

\begin{document}

\twocolumn[\hsize\textwidth\columnwidth\hsize\csname
@twocolumnfalse\endcsname

\title{Novel critical field in magneto-resistance oscillation of 2DEG
in asymmetric GaAs/Al$_{0.3}$Ga$_{0.7}$As double wells measured 
as a function of the in-plane magnetic field}

\draft

\author{P.\ Svoboda$^1$, Y.\ Krupko$^1$, L.\ Smr\v cka$^1$, M.\ Cukr$^1$,
T. Jungwirth$^{1,2}$, and L.\ Jansen$^3$} 

\address{$^1$Institute of Physics ASCR, Cukrovarnick\'a 10, 162 53 Praha
6, Czech Republic\\
$^2$Department of Physics, University of Texas, Austin, TX 78712,  USA\\
$^3$ Grenoble High Magnetic Field Laboratory,
Bo\^{\i}te Postale 166 , 38042 Grenoble Cedex 09, France}

\date{\today}
\maketitle

\begin{abstract}
We have investigated the magnetoresistance of strongly asymmetric
double-well structures formed by a thin $\rm Al_{0.3}Ga_{0.7}As$
barrier grown far from the interface in the GaAs buffer of standard
heterostructures. In magnetic fields oriented parallel to the electron
layers, the magnetoresistance exhibits an oscillation associated with
the depopulation of the higher occupied subband and with the
field-induced transition into a decoupled bilayer.  In addition, the
increasing field transfers electrons from the triangular to
rectangular well and, at high enough field value, the triangular well
is emptied. Consequently, the electronic system becomes a single
layer which leads to a sharp step in the density of electron states
and to an additional minimum in the magnetoresistance curve.
\end{abstract}


\vskip2pc]
\section{Introduction}
A magnetic field $B_{\|}$ applied parallel to the
quasi-two-dimensional systems of electrons confined in double-well
structures is known to couple strongly to the electron orbital motion
and to change dramatically the electron energy spectra. The
magnetoresistance oscillation observed on coupled
double\cite{si,ku,ju,mak} quantum wells represents a striking
manifestation of the two distinct van Hove singularities in the
$B_{\parallel}$-dependent density of states, corresponding to the
depopulation of the antibonding subband at a critical field $B_{||} =
B_{c,1}$, and to the splitting of the Fermi see into two separated
electron sheets at a second critical field $B_{c,2}$.

The structures investigated here were prepared by inserting a thin
Al$_{0.3}$Ga$_{0.7}$As barrier (of a thickness $D$) into the GaAs
buffer layer of a standard modulation-doped
GaAs/Al$_{0.3}$Ga$_{0.7}$As heterostructure, in a distance $d$ from
the interface. The resulting double-well system consists of a nearly
rectangular well of a width $d$ and of a triangular well, both coupled
through a thin barrier. With a proper choice of growth parameters, one
can control the occupancy of the two wells and simultaneously tune the
position of two lowest energy subbands (bonding and
antibonding). Provided that the majority of electrons is in the
bonding subband, it is possible to study the transfer of electrons
between the subbands and/or across the barrier separating both wells
under the influence of the in-plane magnetic field.

\begin{figure}[bt]
\includegraphics[angle=-90,width=7.5cm]{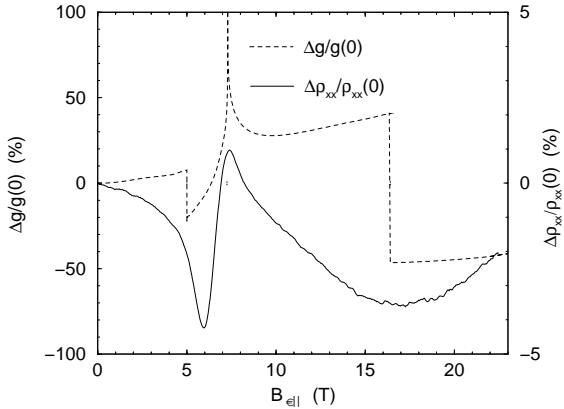}
\caption{Changes induced by parallel magnetic field $B_{\|}$ in DOS,
$\Delta g/g$, and longitudinal resistivity $\Delta
\rho_{xx}/\rho_{xx}$.}
\label{fig1}
\end{figure}

Unlike bilayers realized in conventional double wells or wide single
wells, our system is highly asymmetric. At $B_{\|} = 0$, the bonding
subband electrons, with concentration $n_b$, have a dominant weight in
the rectangular well, while the electrons from the antibonding
subband, with concentration $n_a$ , are more likely to occupy the
triangular well. In our samples $n_b> n_a$ and the corresponding Fermi
contours are two concentric circles.  Applying the in-plane field
$B_{\|}$ induces a transfer of antibonding electrons to the bonding
subband and, therefore, also from the triangular well into the
rectangular one \cite{mak}. The Fermi contour of the antibonding
subband changes its shape and shrinks; at a critical field $B_{c,1}$
it disappears and all electrons occupy only the lowest, bonding
subband. The system behaves as a wide single-layer with electrons
distributed between the two coupled quantum wells. The
magnetoresistance approaches a pronounced minimum (see
Fig.~\ref{fig1}).

Upon further increasing $B_{\|}$, a neck in the peanut-like Fermi
contour narrows which, eventually, results into splitting of the
contour at $B_{\|} = B_{c,2}$.  At this second critical field, the
system undergoes a transition from a single-layer to a bilayer state,
corresponding to a sharp maximum on the magnetoresistance curve (see
Fig.~\ref{fig1}).  Above $B_{c,2}$, electrons of the larger contour
have a dominant weight in the rectangular well, while the smaller
contour electrons occupy the triangular well.  The asymmetry of charge
distribution between the two wells increases with $B_{\|}$.  At a
third critical field $B_{c,3}$ all electrons are transfered into the
rectangular well and the 2D system re-enters the single-layer
state. It is the aim of this paper to study, if the corresponding van
Hove singularity in the DOS can be detected on the magnetoresistance
curve as well.

In the following sections we establish a correspondence between the
van Hove singularity in the DOS at $B_{c,3}$ and a minimum on a
measured magnetoresistance curve.
\section{Experiments}
The growth parameters of our GaAs/Al$_{0.3}$Ga$_{0.7}$As have been
designed to allow for the detection of expected anomalies within the
experimentally accessible range of in-plane magnetic fields. The
structure described here has been grown by MBE with $d$ = 39
monolayers (11.04 nm) and $D$ = 9 monolayers (2.55 nm).  These
parameters are fitted to give the best agreement between the critical
fields $B_{c,1}$, $B_{c,2}$ determined from experiments and the
self-consistent field calculations \cite{ju} for measured $n$, $n_b$
and $n_a$. The $d$ is larger then its nominal value by approximately
10\%, the nominal value of $D$ was 12 monolayers.
The magnetoresistance has been recorded at $T$ = 0.4 K in magnetic
fields up to 23 T. The sample of a standard Hall bar geometry has been
mounted into a rotating sample holder, allowing to adjust the angle
$\alpha$ between $B$ and 2DEG plane to any value between the
perpendicular ($\alpha$ = 90$^{\circ}$) and parallel ($\alpha$ =
0$^{\circ}$) configurations. In all cases, however, the in-plane of
$B$ was perpendicular to the measuring current.  A standard
a.c. lock-in technique ($f$ = 13 Hz ) has been used to simultaneously
determine both longitudinal ($\rho_{xx}$) and Hall ($\rho_H$)
resistivities as a function of applied magnetic field. The low-field
slope of the Hall resistivity, compared to the slope at $\alpha$ =
90$^{\circ}$, has been used to calculate the tilt angle.

The basic characteristics of 2DEG have been extracted from the  data
taken in perpendicular magnetic fields. The total electron
concentration $n = 3.3\times 10^{15}$ m$^{-2}$ was from the low-field
Hall resistivity. Measured $\rho_H$ and $\rho_{xx}$, give the electron
mobility $\mu_H$ = 11.6\,~T$^{-1}$. While the occupancy of two
subbands was evident from the Shubnikov-de Haas oscillations, the
concentration of electrons in the second (antibonding) subband was
apparently very small, causing only a weak modulation of the
oscillations originating from the bonding subband. Fourier analysis
could therefore reliably provide only the concentration of the
bonding electrons $n_b = 3.0 \times 10^{15}$ m$^{-2}$. It implies,
that only about 9\,~\% of all available carriers resides in the
antibonding subband, i.e., in the triangular well in a zero in plane
field.
\begin{figure}[hbt]
\includegraphics[angle=-90,width=7.5cm]{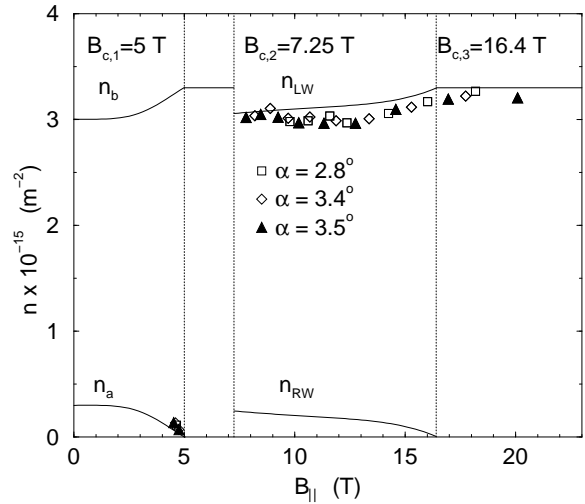}
\caption{Parallel field dependence of the electron concentrations in
the two occupied subbands, $n_b$ and $n_a$. Electrons can be
assigned to individual wells for $B_{\|} >B_{c,2}$, $n_{LW}$
corresponds to the left well and $n_{RW}$ to the right well.}  
\label{fig2}
\end{figure}
%
\section{Results and discussion}
Longitudinal magnetoresistance of the sample adjusted to nearly
parallel configuration ($|\alpha | < 0.1^{\circ}$) is presented in
Fig.~\ref{fig1}, together with the calculated field dependence of the
density of states. In addition to the van Hove singularities related
to $B_{c,1}$ and $B_{c,2}$, there is another minimum on the
magnetoresistance curve that can be associated with the third
singularity of the density of states at $B_{\|} = B_{c,3}$. Recall
that $B_{c,3}$ corresponds to depletion of the triangular well. For
this particular sample we got $B_{c,3}$ = 16.4\,~T.  Fig.~\ref{fig2}
shows the evolution of electron concentrations in the occupied
subbands and/or wells as a function of $B_{\|}$. Full lines represent
results of the theoretical calculation, experimental points have been
determined from the $\rho_{xx}(B)$ curves measured in magnetic fields
slightly tilted from the 2D plane ($\alpha < 4^{\circ}$). Low tilt
angles guarantee, that $B_{\perp} << B_{\|}$, which justifies the
approximations used in the calculation \cite{mak}.  The perpendicular
component of the field induces Shubnikov-de Haas (SdH) oscillations
and makes it possible to estimate the 2D electron densities.  In
lowest magnetic fields, $B_{||} < B_{c,1}$, only the oscillations
arising from antibonding electrons can be seen, which is due to their
lower effective mass. Their concentration is, however, very low and
only few peaks could be identified on all measured curves. The value
of $B_{c,1}$ = 5.0 T, defined by extrapolation $n_a \to 0$, is
therefore less certain. Unlike samples with narrower barriers measured
in Ref.~\cite{mak}, we were not able to reliably detect any
oscillations, that could be attributed to the "single-layer" regime at
$B_{c,1} < B_{\|} < B_{c,2}$, since here the interval of fields is
much narrower.

The oscillations observed at high parallel fields ($B_{\|} > B_{c,2}$
= 7.25 T) arise from two separated Fermi seas of electrons with
concentrations $n_{RW}$ and $n_{LW}$, localized in the right
(triangular) and left (rectangular) wells, respectively. Since
$n_{LW}~\gg~n_{RW}$ and electrons in the triangular well have much
larger effective mass, dominant SdH oscillations arise from the left
well and the other electrons are responsible only for a weak
modulation of the SdH data. Since both concentrations are
field-dependent, the observed oscillations are not periodic in
$1/B_{\perp}$ and Fourier analysis cannot be employed to determine the
periods. Instead, following procedure has been used to estimate
$n_{LW}(B_{\|})$ from the curves measured at low tilt angles.

We assume that the concentration $n_{LW}(B_{\|})$ does not change
within one SdH peak. From the known value of $\alpha$ we find
$B_{\perp}$ and convert the data to the $\rho_{xx}$ vs $1/B_{\perp}$
dependence. We identify the distance between any two subsequent minima
with the period in $1/B_{\perp}$, calculate corresponding electron
concentration and assign it to the position of particular SdH peak on
the $B_{\|}$ axis. This procedure gave the experimental points shown
in Fig.~\ref{fig2} for three different tilt angles. The discrepancy
between data points corresponding to different tilt angles is due to
the fact, that SdH peaks are markedly asymmetric and therefore their
width cannot be determined unambiguously on the $1/B_{\perp}$
scale. The increase of $n_{LW}$ as $B_{\|}$ approaches $B_{c,3}$ is,
however, evident.
\begin{figure}[hbt]
\includegraphics[angle=-90, width=7.5cm]{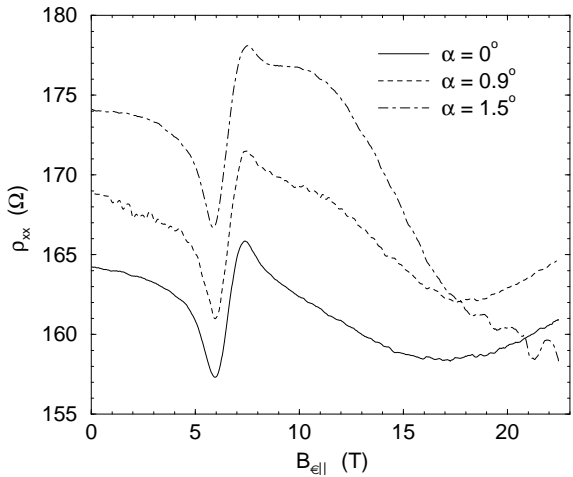}
\caption{Longitudinal magnetoresistance at smallest tilt angles $\alpha$.
For clarity, middle and uppermost curves have been shifted upwards by 
5 $\Omega$ and 10 $\Omega$, respectively.}
\label{fig3}
\end{figure}

The behavior of the sample above $B_{c,2}$ is extremely sensitive to
the tilt angle. This is illustrated in Fig.\,~\ref{fig3}, where we
compare curves for two smallest measured tilt angles with that
obtained at $\alpha =0^{\circ}$.  For clarity, the upper two curves in
the graph have been shifted upwards by 5 $\Omega$ and 10 $\Omega$,
respectively. At $\alpha > 1^{\circ}$, the minimum accompanying
$B_{c,3}$ disappeared.
\section{Conclusions}
We have reported on a novel anomaly in the magnetoresistance of an
asymmetric double quantum well system exposed to strong magnetic
fields parallel to the GaAs/AlGaAs interface.  A minimum is observed
on the $\rho_{xx}(B_{\|}$) curve positioned at a new critical field
$B_{c,3}$. This field agrees quantitatively with the calculated
singularity in the DOS of the DQW, that accompanies a depletion of the
triangular well. Above $B_{c,3}$, electron system behaves as a single
layer 2DEG.

Unlike the lower critical fields $B_{c,1}$ and $B_{c,2}$, that have
been observed and explained before \cite{mak}, the novel critical
field is much more sensitive to a proper choice of the growth
parameters. It can be observed in experimentally accessible magnetic
fields only if the DQW is highly asymmetric in the sense, that the
population of the higher, antibonding subband is only a small fraction
of the overall electron concentration.
\section*{Acknowledgements}
This work has been supported by the Grant Agency of the Czech Republic
under Grant No. 202/01/0754, by the French-Czech project Barrande
99011, by the European Community project ``Access to Research
Infrastructure action of the Improving Human Potential Programme'',
and by the Welch Foundation.

\end{document}